\documentclass{emulateapj}

\newcommand{\myemail}{cmorgan@usna.edu}

\submitted{Submitted to {\it The Astrophysical Journal Letters}}

\shorttitle{QUASAR ACCRETION DISK SIZES}
\shortauthors{MORGAN ET AL.}

\begin{document}

\title{The Quasar Accretion Disk Size -- Black Hole Mass Relation\footnote{B\MakeLowercase{ased on 
observations obtained with the} S\MakeLowercase{mall and} M\MakeLowercase{oderate} A\MakeLowercase{perture} R\MakeLowercase{esearch} T\MakeLowercase{elescope}
S\MakeLowercase{ystem} (SMARTS) 1.3\MakeLowercase{m, which is operated by the} SMARTS C\MakeLowercase{onsortium,
the} A\MakeLowercase{pache} P\MakeLowercase{oint} O\MakeLowercase{bservatory 3.5m telescope, which is 
owned and operated by the} A\MakeLowercase{strophysical} R\MakeLowercase{esearch} C\MakeLowercase{onsortium, the} WIYN O\MakeLowercase{bservatory
which is owned and operated by the} U\MakeLowercase{niversity of} W\MakeLowercase{isconsin}, I\MakeLowercase{ndiana} U\MakeLowercase{niversity}, 
Y\MakeLowercase{ale} U\MakeLowercase{niversity and the} N\MakeLowercase{ational} O\MakeLowercase{ptical} A\MakeLowercase{stronomy} O\MakeLowercase{bservatories} (NOAO), \MakeLowercase{the 
6.5m} M\MakeLowercase{agellan} B\MakeLowercase{aade telescope, which is a collaboration between the observatories 
of the} C\MakeLowercase{arnegie} I\MakeLowercase{nstitution of} W\MakeLowercase{ashington} (OCIW), U\MakeLowercase{niversity of} A\MakeLowercase{rizona}, 
H\MakeLowercase{arvard} U\MakeLowercase{niversity}, U\MakeLowercase{niversity of} M\MakeLowercase{ichigan, and} M\MakeLowercase{assachusetts} I\MakeLowercase{nstitute 
of} T\MakeLowercase{echnology, and observations made with the} NASA/ESA H\MakeLowercase{ubble} S\MakeLowercase{pace} T\MakeLowercase{elescope
for program} HST-GO-9744 \MakeLowercase{of the} S\MakeLowercase{pace} T\MakeLowercase{elescope} S\MakeLowercase{cience} I\MakeLowercase{nstitute,
which is operated by the} A\MakeLowercase{ssociation of} U\MakeLowercase{niversities for} R\MakeLowercase{esearch
in} A\MakeLowercase{stronomy}, I\MakeLowercase{nc., under} NASA \MakeLowercase{contract} NAS 5-26555.
}}

\author{Christopher W. Morgan\altaffilmark{1}}
\affil{Department of Physics, United States Naval Academy, 572C Holloway Road,
Annapolis, MD 21402; \myemail}
 
\author{C.S. Kochanek and Nicholas D. Morgan}
\affil{Department of Astronomy, The Ohio State University, 140 West 18th Avenue, Columbus, OH 43210
-1173; ckochanek@astronomy.ohio-state.edu, nmorgan@astronomy.ohio-state.edu}

\and

\author{Emilio E. Falco}
\affil{Harvard-Smithsonian Center for Astrophysics, 60 Garden Street, Cambridge, MA, 02138; 
efalco@cfa.harvard.edu}

\altaffiltext{1}{Department of Astronomy, The Ohio State University}

\clearpage

\begin{abstract}
We use the microlensing variability observed for nine 
gravitationally lensed quasars to show that the accretion disk size at 2500\AA~is 
related to the black hole mass by 
$\log(R_{2500}/$cm$)=(15.6\pm0.2) + (0.54\pm0.28)\log(M_{BH}/10^9$M$_{\sun})$.  
This scaling is consistent 
with the expectation from thin disk theory ($R \propto M_{BH}^{2/3}$), but it implies that black 
holes radiate with relatively low efficiency, $\log(\eta)  = -1.29\pm0.44 + \log(L/L_E)$
where $\eta=L/(\dot{M}c^2)$.  These 
sizes are also larger, by a factor of $\sim3$, than the size needed to produce the observed 
$0.8\,$\micron~ quasar flux by thermal radiation from a thin disk with the same $T \propto R^{-3/4}$ 
temperature profile.  More sophisticated disk models are clearly required, 
particularly as our continuing observations improve the precision of the 
measurements and yield estimates of the scaling with wavelength and accretion 
rate.

\end{abstract}


\section{Introduction}
\label{sec:introduction}

Despite nearly 40 years of work on accretion disk physics, the simple \citet{Shakura1973} 
thin disk model and its relativistic cousins \citep[e.g.][]{Page1974,Hubeny1997,Li2005}
remain the standard model despite many theoretical 
alternatives \citep[e.g.][]{Narayan1997,DeVilliers2003,Blaes2007} and some observational 
reservations \citep[see][]{Collin2002}. 
Quasar accretion disks cannot be spatially resolved with ordinary telescopes, so we have 
been forced to test accretion physics through time variability \citep[e.g.][]{VandenBerk2004,
Sergeev2005,Cackett2007} and spectral modeling \citep[e.g.][]{Sun1989,Bonning2007}.  
One notable success is the use of reverberation mapping \citep{Peterson2004} of quasar broad 
line emission to calibrate the relation between emission line widths and black hole 
masses. 

Gravitational telescopes do, however, provide the necessary resolution to study the 
structure of the quasar continuum source.  Each gravitationally lensed quasar image is observed through a 
magnifying screen created by the stars in the lens galaxy.  Sources that are smaller than the Einstein 
radius of the stars, typically $\sim10^{16}$ cm, show time variable fluxes whose amplitude is 
determined by the source size \citep[see the review by][]{Wambsganss2006}. 
Smaller sources have larger variability amplitudes than 
larger sources. In this investigation, we exploit the optical microlensing 
variability in nine gravitationally lensed quasar systems to measure the size of their accretion disks,
and we find that disk sizes are strongly correlated with the masses of their central black holes. 

\begin{deluxetable*}{lccccc}[b]
\tablewidth{0pt}
\tablecaption{Derived Quantities}
\tablehead{\colhead{Object} 
		& \colhead{$M_{BH}$} 
		& \colhead{$\log(R_S/{\rm cm})$}
		& \colhead{$\lambda_{rest}$}
		& \colhead{$I_{corr}$}
		& \colhead{$\log(R_S/{\rm cm})$} \\
		\colhead{}
		& \colhead{$(10^9~{\rm M_\sun})$}
		& \colhead{(microlensing)}
		& \colhead{(\micron)}
		& \colhead{(mag)}
		& \colhead{(thin disk flux)}
		}		
\startdata
HE0435-1223 & 0.50 & $15.6_{-0.7}^{+0.5}$ & 0.260 & $20.76\pm 0.25$ & $14.9\pm0.1$ \\
SDSS0924+0219 & 0.11 & $14.8_{-0.4}^{+0.3}$ & 0.277 & $21.24\pm 0.25$ & $14.8\pm0.1$  \\
FBQ0951+2635 & 0.89 & $15.9_{-0.4}^{+0.4}$ & 0.313 & $17.16\pm 0.11$ & $15.6\pm0.1$ \\
HE1104-1805 & 2.37 & $15.7_{-0.3}^{+0.2}$ & 0.211 & $18.17\pm 0.31$ & $15.4\pm0.1$ \\
PG1115+080 & 0.92 & $16.5_{-0.9}^{+0.5}$ & 0.257 & $19.52\pm 0.27$ & $15.1\pm0.1$ \\
RXJ1131-1231 & 0.06 & $15.4_{-0.2}^{+0.3}$ & 0.422 & $20.73\pm 0.11$ & $14.8\pm0.1$ \\
SDSS1138+0314 & 0.04 & $14.8_{-0.6}^{+0.6}$ & 0.203 & $21.97\pm 0.19$ & $14.6\pm0.1$ \\
SBS1520+530 & 0.88 & $15.5_{-0.2}^{+0.2}$ & 0.245 & $18.92\pm 0.13$ & $15.3\pm0.1$ \\
Q2237+030 & 1.3~ & $15.5_{-0.3}^{+0.3}$ & 0.208 & $18.03\pm 0.44$ & $15.5\pm0.2$ \\
\enddata
\tablecomments{$R_S$ from microlensing is the accretion disk size at
$\lambda_{rest}$, the rest-frame wavelength corresponding to the 
center of the monitoring filter used for that quasar's lightcurve. 
$I_{corr}$ is the corrected (unmagnified) $I$-band magnitude.
Typical $I$-band measurement errors are $\lesssim 0.1$ mag, but the larger errors
on $I_{corr}$ come from uncertainties in the lens magnification.
$R_S$ calculated using corrected $I$-band magnitude and thin disk theory is also unscaled; it is the 
disk size at the rest-frame wavelength corresponding to the center of the 
{\it HST} $I$-band filter (F814W).  Both disk sizes assume an average inclination angle $i=60\degr$.}
\label{tab:rs}
\end{deluxetable*}
 
In \S\ref{sec:observations} we describe the monitoring data, the lens models we use based
on Hubble Space Telescope (HST) images of each system and our microlensing analysis method.
In \S\ref{sec:results} we describe our accretion disk model and our results for the relationship
between disk size and black hole mass. In \S\ref{sec:discussion}, 
we discuss the results and their implications for thin accretion disk theory.
All calculations in this paper
assume a flat $\Lambda$CDM cosmology with $h=0.7$, $\Omega_{M}=0.3$ and $\Omega_{\Lambda}=0.7$.

\section{Data and Analysis}
\label{sec:observations}

We monitored the gravitationally lensed quasars HE 0435--1223, SDSS 0924+0219, FBQ 0951+2635, HE 1104--1805, 
PG 1115+080, RXJ 1131--1231, SDSS 1138+0314, SBS 1520+530 and Q 2237+030 
in the $R$- and $V$-bands on the SMARTS 1.3m using 
the ANDICAM optical/infrared camera \citep{Depoy2003}\footnote{http://www.astronomy.ohio-state.edu/ANDICAM/},
the Wisconsin-Yale-Indiana (WIYN) observatory
using the WIYN Tip--Tilt Module (WTTM) \footnote{http://www.wiyn.org/wttm/WTTM\_manual.html},
the 2.4m telescope at the MDM Observatory using the MDM Eight-K
\footnote{http://www.astro.columbia.edu/~arlin/MDM8K/}, Echelle and RETROCAM
\footnote{http://www.astronomy.ohio-state.edu/MDM/RETROCAM} \citep{Morgan2005}
imagers and the 6.5m Magellan Baade telescope using IMACS \citep{Bigelow1999}.
We supplemented our monitoring data with published quasar light curves  
from \citet{Paraficz2006}, \citet{Schechter1997}, \citet{Ogle2003},  
\citet{Ofek2003} and \citet{Gaynullina2005}.  
We measured the flux of each image by comparison to the flux from reference stars 
in the field of each frame. Our analysis of the monitoring data is described
in detail by \citet{Kochanek2006}. 

\begin{figure*}[t]
\epsscale{1.0}
\plotone{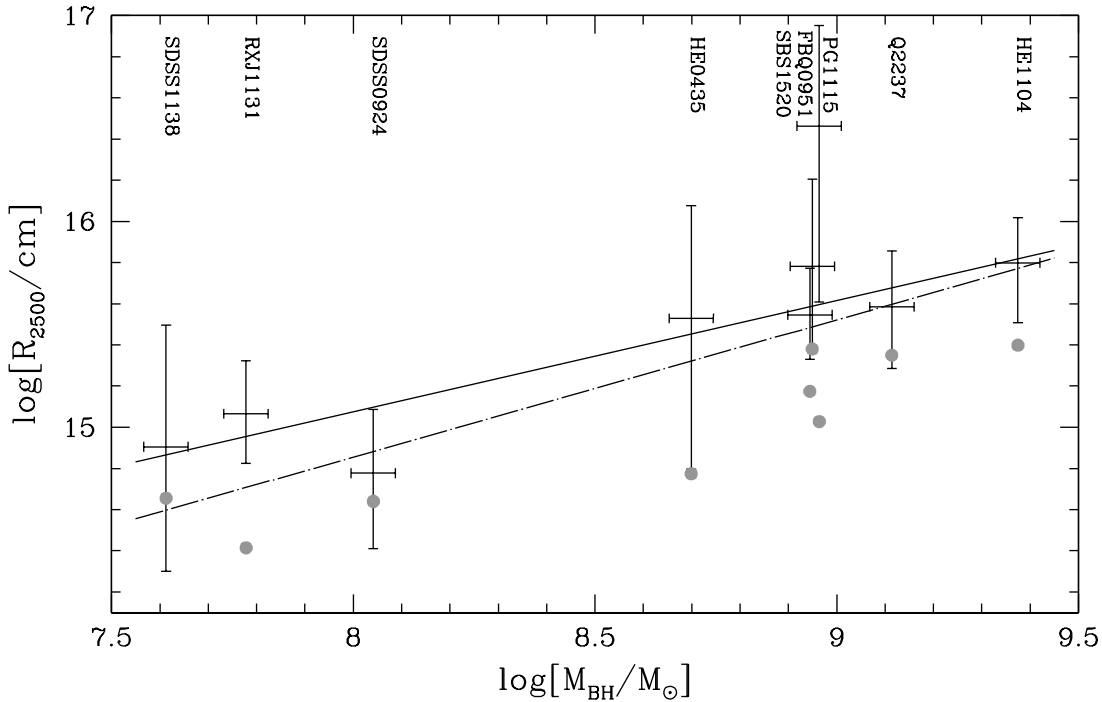}
\caption{Inclination-corrected accretion disk size $R_{2500}$  versus 
black hole mass $M_{BH}$. The solid line shows our best power-law fit to the data 
and the dot-dashed line shows the prediction from thin disk theory ($L/L_E = 1$ and 
$\eta = 0.1$).  Disk sizes are corrected to a rest wavelength of  $\lambda_{rest} = 2500$\AA~and the 
black hole masses were estimated using emission line widths.  The filled points 
without error bars are $R_{2500}$ estimates based on the observed, magnification-corrected 
$I$-band fluxes.  They have typical uncertainties of $0.1$-$0.2$~dex.
\label{fig:r2500}}
\end{figure*}

All nine lenses have been observed in the $V$- (F555W), $I$- (F814W) and $H$-bands (F160W) using
the WFPC2, ACS/WFC and NICMOS instruments on {\it HST}.  We fit these images as combinations of
point sources for the quasars and (generally) de Vaucouleurs models for the lenses as
described in  \citep{Lehar2000}.  These provided the astrometry used for lens models
and defined a constant mass-to-light ($M/L$) ratio model for the mass distribution in the 
lens models.  We modeled each system using the {\it GRAVLENS} software package \citep{GRAVLENS} 
to generate a series of ten models for each lens starting from a constant $M/L$ model
and then adding an NFW \citep{Navarro1996} halo.  The sequence is parametrized by  
$f_{M/L}$, the mass fraction represented by the visible lens galaxy relative to a 
constant $M/L$ model. We start with the constant $M/L$ model, $f_{M/L}=1$, and then
reduce its mass in increments of $\Delta f_{M/L}=0.1$ with the NFW halo's mass rising
to compensate.  

These lens models then provide the convergence $\kappa$, shear $\gamma$ and
stellar surface density $\kappa_*$ needed to define the microlensing magnification
patterns.  We assume a lens galaxy stellar mass function $dN(M)/dM \propto M^{-1.3}$ with a 
dynamic range of a factor of 50 that approximates the Galactic disk mass function 
of \citet{Gould2000}.  For the typical lens we generated 4 magnification 
patterns for each image in each of the 10 lens models. We gave 
the magnification patterns an outer scale of $20 \langle R_E\rangle$, where 
$\langle R_E \rangle$ is the Einstein radius for the mean stellar mass $\langle M\rangle$.
This outer dimension is large enough to fairly sample the magnification pattern while
the pixel scale is small enough to resolve the accretion disk.
We determined the properties of the accretion disk by modeling the observed light curves
using the Bayesian Monte Carlo method of \citet{Kochanek2004} \citep[also see][]{Kochanek2007}.  
For a given disk model we randomly generate light curves, fit them to the observations 
and then use Bayesian methods to compute probability distributions for the disk size
averaged over the lens models, the likely velocities of the observer, lens, source, stars, 
and mass. We use a prior on the mean microlens mass of $0.1 {\rm M_\sun} < \langle M \rangle < 1.0{\rm M_\sun}$, 
but the disk size estimates are relatively insensitive to this assumption \citep[see][]{Kochanek2004}.
 
We use black hole mass estimates for the quasars that are based on observed quasar emission 
line widths and the locally calibrated virial relations for black hole masses \citep{Onken2004,Greene2007}.  
For most systems we simply adopted the black hole mass estimates from \citet{Peng2006} based on
the \ion{C}{4}~($\lambda 1549$\AA), \ion{Mg}{2}~($\lambda 2798$\AA) and H$\beta$~($\lambda 4861$\AA) 
mass-linewidth relations.  For SDSS 1138+0314, we measured the width of the \ion{C}{4}
(1549\AA) line in optical spectra from the Sloan Digital Sky Survey \citep{SDSSDR4} and estimated the 
black hole mass using the normalizations of \citet{Vestergaard2006}.  These mass estimates
are reliable to approximately 0.3~dex \citep[see][]{McLure2002,Kollmeier2006,Vestergaard2006,Peng2006}. 

\section{Results}
\label{sec:results} 

We model the surface brightness profile of the accretion disk as a power law 
temperature profile, $T \propto R^{-3/4}$, matching the outer regions of a \citet{Shakura1973} thin 
disk model.  We neglect the central depression of the temperature due to the inner edge 
of the disk and corrections from general relativity to avoid extra parameters.  The effect 
of this simplification on our size estimates is small compared to our measurement 
uncertainties provided the disk size we obtain is several times larger than the 
radius of the inner disk edge.  We assume that the disk radiates as a black body, so the surface 
brightness at rest wavelength $\lambda_{rest}$ is
\begin{equation}
f_{\nu} = {2 h_p c \over \lambda_{rest}^3} 
\left[ \exp \left( {R \over R_{\lambda_{rest}}} \right)^{3/4}-1 \right]^{-1}
\end{equation}                                                                                   
where the scale length
\begin{eqnarray}
R_{\lambda_{rest}} & = & \left[ {45 G \lambda_{rest}^4 M_{BH} \dot{M} \over 16 \pi^6 h_p c^2} \right]^{1/3} \nonumber \\
& = & 9.7 \times  10^{15}  \left( {\lambda_{rest} \over {\rm \micron}} \right)^{4/3}\nonumber  \\
& & \times \left( {M_{BH} \over 10^9 {\rm M_{\sun}}} \right)^{2/3}
\left({L \over \eta L_E} \right)^{1/3} {\rm cm}
\end{eqnarray}
is the radius at which the disk temperature matches the wavelength, 
$k T_{\lambda_{rest}} = h_p c/ \lambda_{rest}$, 
$h_p$ is the Planck constant, $k$ is the Boltzmann constant, $M_{BH}$ is the black hole mass,   
$\dot{M}$ is the mass accretion rate, $L/L_E$ 
is the luminosity in units of the Eddington luminosity, and $\eta=L/(\dot{M}c^2)$ is the accretion efficiency.   
We can also compute the size under the same model assumptions based on the 
magnification-corrected $I$-band quasar fluxes measured in HST observations as
\begin{eqnarray}
R_I  &=& 2.83 \times 10^{15} {1 \over \sqrt{\cos i}}
\left( { D_{OS} \over r_H } \right) \nonumber \\
&& \times \left( { \lambda_{I,obs} \over {\rm \micron} } \right)^{3/2}
10^{-0.2(I-19)} \: h^{-1} \: {\rm cm}
\end{eqnarray}
where $D_{OS}/r_H$ is the angular diameter distance to the quasar in units of the Hubble 
radius, $I$ is the magnification-corrected magnitude and $i$ is the disk inclination angle.

\begin{figure}[t]
\epsscale{1.0}
\plotone{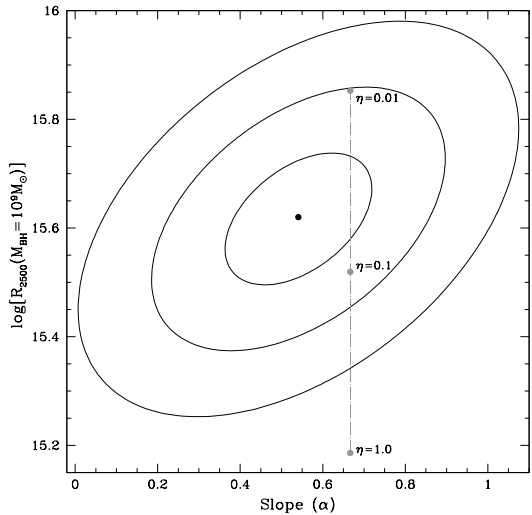}
\caption{Results of the power-law fit to $R_{2500}$ as a function of black 
hole mass.  The contours show the $1-3\sigma$ one-parameter confidence 
intervals for the slope $\alpha$ 
and the normalization $R_{2500}(M_{BH} = 10^9 {\rm M_{\sun}})$ for the 2500\AA~accretion disk size 
corresponding to $M_{BH} = 10^9 {\rm M_{\sun}}$. The best-fit value is indicated with a black point. 
The filled points along the dot-dashed line are theoretical thin disk sizes for quasars radiating at the 
Eddington limit and with efficiencies of $\eta = L / (\dot{M} c^2)  = 0.01$, $0.1$ or $1.0$.
\label{fig:r2500fit}}
\end{figure}

\begin{figure}[t]
\epsscale{1.0}
\plotone{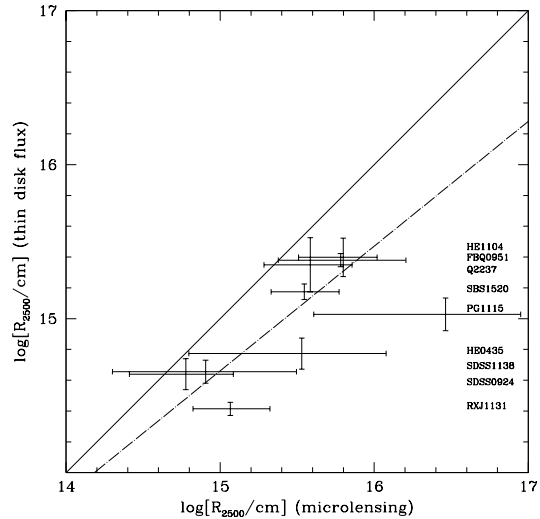}
\figcaption{Thin disk flux size estimates versus accretion disk sizes 
from microlensing.  For reference, the solid line indicates a one-to-one 
relationship between thin disk flux size estimates and microlensing 
measurements.  The dot-dashed line is the best fit to the data. Since the data 
points have large errors relative to their dynamic range, the best-fit slope is 
consistent with unity and its average offset from the solid line is $0.4$~dex.
\label{fig:rflux}}
\end{figure}

Our results are shown in Figures~\ref{fig:r2500} through~\ref{fig:rflux} 
and summarized in Table~\ref{tab:rs}.  
For the comparison with theory and the figures, we corrected the measured sizes to 
$\lambda_{rest} = 2500{\rm \AA}$ assuming the $\lambda^{4/3}$  scaling of thin disk theory 
and the mean inclination $\langle \cos i \rangle = 1/2$. 
There are two striking facts illustrated by the figures.  First, we clearly see 
from Fig.~\ref{fig:r2500} that the microlensing sizes are well correlated with the black hole mass.  A 
power-law fit between $R_{2500}$ and $M_{BH}$ yields: 
\begin{eqnarray}
 \log \left( {R_{2500}\over {\rm cm}} \right)&=&(15.6\pm0.2) \nonumber \\
&&+(0.54\pm0.28)\log \left( {M_{BH} \over 10^9{\rm M_{\sun}}} \right)
\end{eqnarray}
which is consistent with the predicted slope from thin disk theory ($R \propto M_{BH}^{2/3}$) and 
implies a typical Eddington factor of $\log(L/\eta L_E)=1.29\pm0.44$ if we fix the slope with 
mass to 2/3 (see Fig. 2).  \citet{Kollmeier2006} estimate that the typical quasar has $L/L_E \approx 1/3$, 
which would indicate a radiative efficiency of $\eta = L / (\dot{M} c^2) \simeq 0.02$. This efficiency is 
low compared to standard models \citep[e.g.][]{Gammie1999}.  Second, we find that microlensing sizes are well 
correlated with sizes estimated from thin disk theory and from the observed flux (Figs.~\ref{fig:r2500}
and~\ref{fig:rflux}), but the three size estimates show systematic offsets in scale. Most of the offset 
between the microlensing disk size measurements and thin disk theory size estimates could 
be explained by the existing uncertainties, but the offset from the estimate based on the 
quasar flux is much more significant (Fig.~\ref{fig:r2500}). While the microlensing and
flux sizes are well-correlated, the measured disk sizes are $0.4\pm0.3$~dex 
larger than predicted from the observed flux.  Simply put, the quasars are not 
sufficiently luminous to be radiating as black bodies with a $T \propto R^{-3/4}$ temperature profile. 
\citet{Pooley2006} also noticed this problem in their more qualitative study of lensed quasars 
with X-ray observations.

\section{Discussion}
\label{sec:discussion}

Are the discrepancies between our measurements and the size estimated 
from the disk flux due to a problem in the measurements, an oversimplification of the 
disk model or a fundamental problem in the thin disk model?  We have tested our 
approach using Monte Carlo simulations of light curves and verified that we recover the 
input disk sizes.  Our results are also only weakly sensitive to the assumed prior on the 
microlens masses \citep[see][for a discussion]{Kochanek2004}.  We will overestimate the source 
size if a significant fraction of the observed flux comes not from the continuum 
emission of the disk but from the larger and minimally microlensed line emitting 
regions \citep[e.g.][]{Sugai2007}.  This includes not only the obvious broad lines but also the 
broad \ion{Fe}{2} and Balmer continuum emission that can represent $\sim30\%$ of the apparent 
continuum flux at some wavelengths \citep{Netzer1983,Grandi1982}.  
We have experimented with adding a fraction 
of unmicrolensed light and found that 30\% contamination does not lead to sufficiently large size 
changes to resolve the problem.  The sizes shrink by approximately 20\%. 
Conservatively, black hole mass estimates from the virial technique have a scatter of a 
factor of $\sim3$ \citep{McLure2002,Kollmeier2006,Vestergaard2006,Peng2006}, 
which contributes only 0.3 dex of scatter to the disk size 
estimates.  The size estimates from the flux could be affected by misestimating the 
magnification or failing to correct for extinction in the lens galaxy, but the uncertainties 
in the magnifications are only a factor of $\sim2$ at worst and none of these lenses shows 
significant extinction \citep{Falco1999,Eliasdottir2006} 
relative to the magnitude of the discrepancy.

The problem is also not a consequence of the obvious flaws in our simplified disk model -- 
the neglect of the inner edge and the different temperature 
profile of a relativistic disk \citep[e.g.][]{Page1974}.  
For observations at a fixed wavelength, neglecting the inner edge does not have a 
dramatic effect on the effective source size and the effects of relativity on the 
temperature profile are modest.   Studies of the temperature profile in the disk with 
microlensing are best done by measuring the variation in source size with wavelength 
because size ratios can be measured much more accurately than absolute sizes.
\citet{Poindexter2007} have used the wavelength dependence of the microlensing
in HE 1104-1805 to derive a slope $T \propto R^{-\beta}$ or $R_\lambda \propto \lambda^{1/\beta}$
of $\beta = 0.61_{-0.17}^{+0.21}$ that is consistent with thin disk theory but would also allow a 
shallower temperature profile that would reduce the differences between the microlensing 
and flux size estimates.   
The next step is clearly to use more sophisticated disk models including 
relativity and model atmospheres such as \citet{Hubeny1997} or \citet{Li2005}. \citet{Mortonson2005} 
have argued that microlensing essentially measures the half-light radius ($R_{1/2} = 
2.44R_{\lambda}$ for our model), so the first step should be to try to simultaneously match our size 
estimates and the observed fluxes assuming this to be the case, since the many 
additional parameters of the full disk models will make their direct inclusion in the 
microlensing calculations a major computational challenge. 
 
\acknowledgments
We thank E. Agol, M. Dietrich, C. Onken, B. Peterson, M. Pinsonneault, R. Pogge and 
P. Osmer for discussions on quasar structure, and M. Mortonson, S. Poindexter, S. Rappaport, and P. 
Schechter for discussions on microlensing. This research made extensive use
of a Beowulf computer cluster obtained through the Cluster Ohio
program of the Ohio Supercomputer Center. Support for program HST-GO-9744 was
provided by NASA through a grant from the Space Telescope Science Institute, which 
is operated by the Association of Universities for Research in Astronomy, Inc., under
NASA contract NAS-5-26666. 

{\it Facilities:} \facility{CTIO:2MASS (ANDICAM)}, \facility{Hiltner (RETROCAM)}, \facility{WIYN (WTTM), 
\facility{HST (NICMOS, ACS)}}.

\end{document}